\documentclass[sigconf, nonacm] {acmart} 
\usepackage{caption}
\usepackage{multirow}
\usepackage{xcolor}
\usepackage{graphicx}
\usepackage{epstopdf}
\usepackage{url}
\usepackage{comment}
\usepackage{graphicx}
\usepackage{caption}
\usepackage{subcaption}
\usepackage{amsmath} 
\usepackage{tabularx}
\AtBeginDocument{%
  \providecommand\BibTeX{{%
  \normalfont 
  \kern-0.5em{\scshape i\kern-0.25em b}\kern-0.8em\TeX}
  }}

\begin{document}

\title{AlphaAgents: Large Language Model based Multi-Agents for Equity Portfolio Constructions}

\author{Tianjiao Zhao}
\email{tina.zhao@blackrock.com}
\affiliation{%
  \institution{BlackRock, Inc.}
  \city{Atlanta, GA}
  \country{USA}
}

\author{Jingrao Lyu}
\email{jingrao.lyu@blackrock.com}
\affiliation{%
  \institution{BlackRock, Inc.}
  \city{Atlanta, GA}
  \country{USA}
}

\author{Stokes Jones}
\email{stokes.jones@blackrock.com}
\affiliation{%
  \institution{BlackRock, Inc.}
  \city{Atlanta, GA}
  \country{USA}
}

\author{Harrison Garber }
\email{harrison.garber@blackrock.com}
\affiliation{%
  \institution{BlackRock, Inc.}
  \city{Atlanta, GA}
  \country{USA}
}

\author{Stefano Pasquali}
\email{stefano.pasquali@blackrock.com}
\affiliation{%
  \institution{BlackRock, Inc.}
  \city{New York, NY}
  \country{USA}
}

\author{Dhagash Mehta}
\email{dhagash.mehta@blackrock.com}
\affiliation{%
  \institution{BlackRock, Inc.}
  \city{New York, NY}
  \country{USA}
  }

\renewcommand{\shortauthors}{Zhao et al.}

\begin{abstract}
The field of artificial intelligence (AI) agents is evolving rapidly, driven by the capabilities of Large Language Models (LLMs) to autonomously perform and refine tasks with human-like efficiency and adaptability. In this context, multi-agent collaboration has emerged as a promising approach, enabling multiple AI agents to work together to solve complex challenges. This study investigates the application of role-based multi-agent systems to support stock selection in equity research and portfolio management. We present a comprehensive analysis performed by a team of specialized agents and evaluate their stock-picking performance against established benchmarks under varying levels of risk tolerance. Furthermore, we examine the advantages and limitations of employing multi-agent frameworks in equity analysis, offering critical insights into their practical efficacy and implementation challenges.

\end{abstract}

\maketitle

\section{Introduction}
Equity portfolio management has traditionally relied on human research analysts for guidance on portfolio construction and monitoring. This involves gathering and processing substantial amounts of information related to the current performance and anticipated future performance of individual equities within a portfolio. Analysts must actively collect, analyze, and synthesize this information to form well-informed judgments. Typical inputs for equity research include financial disclosures (e.g., 10-K and 10-Q reports), earnings calls, price targets, financial ratios, valuation analyses, market news, sector research reports, and notes from company visits, among others. The need to analyze this vast array of diverse and voluminous information presents significant challenges to analysts in supporting their investment teams to generate alpha for clients, in addition to personal perspectives and prior investment experiences that often influence their interpretation, potentially leading fund managers to overlook alpha opportunities. AI offers an obvious advantage in rapidly processing large volumes of data and extracting actionable insights, ultimately aiding analysts in the pursuit of alpha \cite{nie2024survey}.

Recent advances in LLMs, particularly in multi-agent framework \cite{wu2023autogen}, have significantly expanded AI's ability to handle complex reasoning and decision-making using unstructured data \cite{talebirad2023multi}. AI agents, built on LLMs, now incorporate advanced functionalities such as planning, memory management, and tool usage. In this work, we propose a novel multi-agent investment framework that leverages LLM-powered collaboration and debate to collect and synthesize equity research inputs. By employing a team of specialized agents capable of reasoning over market data and fundamentals, we demonstrate that this collaborative framework can improve the investment process and outcomes.

In addition, we explore how LLM based agentic system can mitigate congnitive biases, a central theme in Behavioral Finance. Rooted in Kahneman and Tversky's Prospect Theory \cite{kahneman2013prospect}, this field has shown that biases like loss aversion and overconfidence often lead to suboptimal investment decisions \cite{baker2010behavioral}. Although previous has recognized these challenges \cite{antony2020behavioral, de1995financial, caparrelli2004herding}, practical solutions remain limited \cite{davies2017practical}. Approaches such as awareness checklists \cite{fromlet2001behavioral} or efficient frontier anchoring \cite{evensky2017applications} often fail to address the unconscious nature of these biases \cite{taffler2010emotional}. We argue that multi-agent LLM systems can help overcome these limitations by complementing human equity research, reducing cognitive biases and simultaneously mitigating AI-specific issues such as hallucination through collaborative task delegation to unbiased AI agents.
\subsection{Related Work}
Early applications of agent-based systems in finance have predominantly relied on Reinforcement Learning (RL) for portfolio management \cite{sutton2018reinforcement, lee2020maps, ma2023multi} which are typically task-specific and operate on structured data rather than being powered by language models.

Recent advances in LLMs \cite{naveed2023comprehensive} have catalyzed the emergence of LLM-based autonomous agents \cite{wang2024survey}. These agents exhibit strong capabilities across domains such as strategic decision-making \cite{li2024more}, reflective reasoning \cite{renze2024self}, external tool usage \cite{li2023api}, memory-based learning \cite{dong2022survey}, and grounded planning \cite{song2023llm}. In particular, ollaborative LLM agents \cite{agashe2023evaluating} have demonstrated effectiveness in coordinating tasks through dialogue and debate.


FinRobot \cite{yang2024finrobot} addresses financial statement analysis through agentic specialization. FinMem \cite{yangyang2023finmem} integrates structured financial data with layered memory, although its effectiveness is constrained by news coverage. MarketSenseAI \cite{fatouros2025marketsenseai} incorporates five specialized agents to parse distinct financial modalities. FinAgent \cite{zhang2024multimodal} further expands capabilities by handling multimodal financial data. Finally, FinVerse \cite{an2024finverse} wraps more than 600 financial APIs inside a crew of retrieval-and-tool-calling agents to answer ad-hoc investor queries.

Despite these advances, the use of multi-agent LLMs for systematic stock selection and portfolio construction remains relatively unexplored. Prior studies often emphasize chain-of-thought prompting but neglect structured agent interaction strategies. Moreover, the role of agent-specific risk tolerance in shaping financial decision outcomes has received little attention.

This work addresses these gaps by introducing a role-based multi-agent system for equity research and screening. It incorporates cooperative and adversarial reasoning, leverages domain-specific tools for each agent, and integrates explicit risk tolerance profiles. The proposed framework offers a novel contribution to multi-agent decision-making in equity portfolio construction.

\section{Methodology}
\subsection{Multi-agent System for Equity Research}
Equity research and portfolio management require the comprehensive collection and analysis of diverse data sources, encompassing both textual and numerical information. Examples include corporate financial disclosures, earnings calls, and quantitative analyses derived from historical stock prices and trading volumes. In this study, we demonstrate the application of a multi-agent framework to enhance equity analysis and stock selection. Specifically, we developed three specialized micro-agents, each endowed with distinct expertise and toolsets. These agents are designed to emulate the specialized roles and responsibilities typically performed by equity analysts within a human portfolio management team.

The three specialized AI agents collaborate to analyze individual stocks and produce comprehensive stock analysis reports. This collaborative approach mitigates human behavioral biases and enhances the quality of reasoning by leveraging diverse domain-specific expertise among the agents. The agents may reach differing conclusions when analyzing the same problem, due to either reasoning divergences or underlying hallucination issues. To address this, the framework incorporates an internal debating mechanism, allowing agents to engage in discussions when their analyses diverge. This process continues until consensus is achieved, thereby improving the multi-agent reasoning capability and reducing hallucinations, as supported by recent research \cite{du2023improving}. 

\begin{figure*}
    \centering
    \includegraphics[width=1\linewidth]{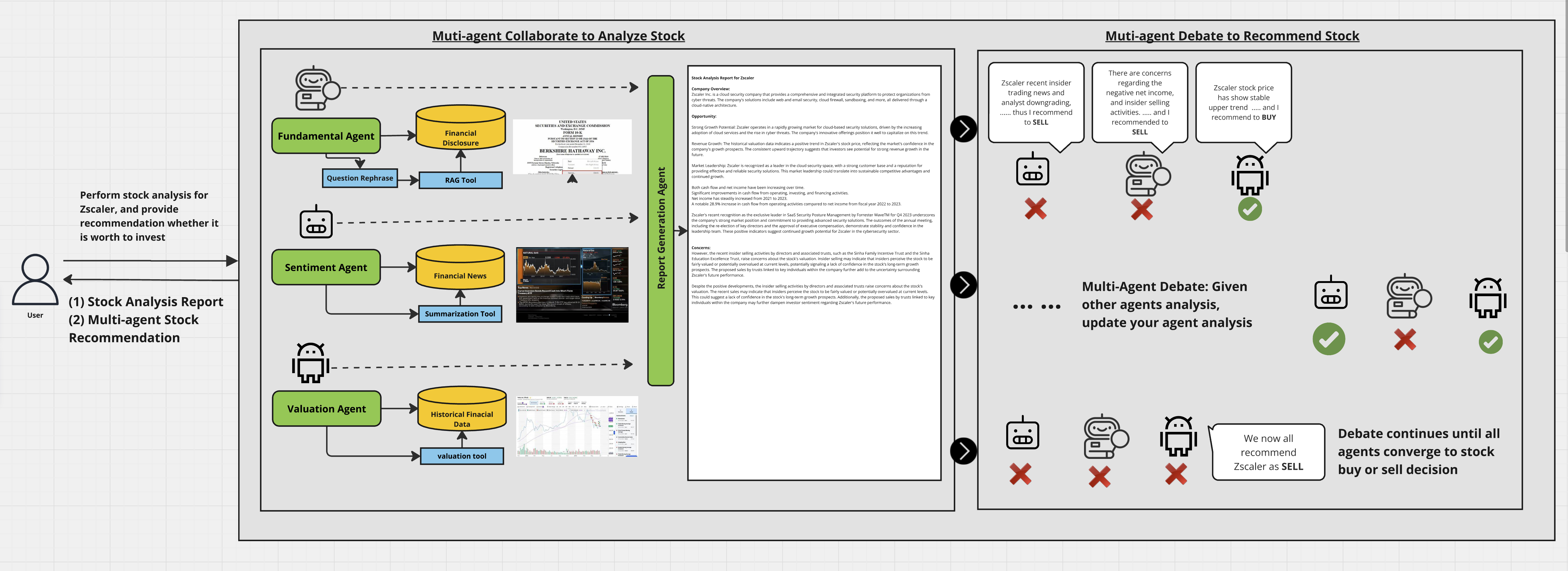}
    \caption{System diagram of multi-agent collaboration and debate for stock analysis}
    \label{fig:workflow}
\end{figure*}

The system diagram presented in Figure \ref{fig:workflow} illustrates the high-level workflow, with the specific roles and responsibilities of each specialist agent detailed below.

\begin{itemize}
    \item \textbf{Fundamental Agent:} Traditionally, fundamental equity analysis has been the cornerstone of equity research, involving in-depth assessments of a stock's value through the examination of 10-K and 10-Q reports, sector and company trends, and financial statement data. Here, we automate this labor-intensive process using an LLM agent capable of qualitatively and quantitatively analyzing an equity's projected trajectory and financial performance.
    \item \textbf{Sentiment Agent:} Financial markets often react to patterns influenced by news or changes in analyst ratings. Such news may pertain to company operating events, recent analyst evaluations, executive changes, or insider trading disclosures. To address this, the Sentiment Agent is designed as a specialist to analyze financial news and provide recommendations based on the prevailing sentiment toward an equity and its potential impact on stock prices.
    \item \textbf{Valuation Agent:} Current and historical prices and volumes capture market expectations of stock performance and offer insights into whether a stock is reasonably priced based on price and volume trends. The Valuation Agent is designed to analyze stock prices and volumes, providing a valuation assessment to inform the stock's relative significance within a portfolio.
\end{itemize}

While this study uses only three agents, the framework is designed for scalability and can easily incorporate additional agents. For example, portfolio managers may benefit from a Technical Analysis Agent for short-term trend detection or a Macro Economist Agent focused on economic policy and macroeconomic data. Future work will explore a more extensive multi-agent system that integrates a broader set of specialized agents.

\subsection{Methodology for Constructing Agents}
\subsubsection{System Data Description}
The primary data sources utilized in the system are listed in Table \ref{tab:data_features}, including 10-K and 10-Q financial disclosures, Bloomberg financial news, and historical stock price and volume data from Yahoo Finance. Each agent is granted access only to the data relevant to its specific role and assigned task.
\begin{table}[htbp]
    \centering
    \begin{tabular}{>{\raggedright\arraybackslash}p{0.8 in} >{\raggedright\arraybackslash}p{0.8 in} >{\raggedright\arraybackslash}p{1.6 in}}
    \toprule
    Feature                  &Availability& Description\\ \midrule
    Ticker &All Agents& Unique symbol of a publicly traded company's stock\\
    Price (Open, High, Low, Close) &Valuation Agent& The initial, highest, lowest, and final trading price of a stock within a specific trading period\\
    Volume &Valuation Agent& The total number of shares traded during a specific period\\
 10K/10Q& Fundamental Agent&Financial disclosure data (includes discussion of major company prospectus and financial statements).\\  
    Bloomberg ID &Sentiment Agent& Ticker's corresponding unique identifier for Bloomberg News\\
    News Body &Sentiment Agent& The main textual content of a news article detailing news, company disclosure, and analyst rating change.\\ 
    \bottomrule
    \end{tabular}
    \caption{Feature description}
    \label{tab:data_features}
    \vspace{-6mm}
\end{table}

\subsubsection{Role Prompting}
Role prompting is a prompt engineering technique \cite{sahoo2024systematic} that involves guiding a conversational agent by assigning it a specific role or context, enabling the LLM to generate responses that are more relevant and contextually appropriate. In our multi-agent setup, we utilized role prompting for the three agents by clearly defining their functions and objectives to enhance task-specific performance and accuracy.
\begin{enumerate}
    \item Valuation Agent: "\textit{As a valuation equity analyst, your primary responsibility is to analyze the valuation trends of a given asset or portfolio over an extended time horizon. To complete the task, you must analyze the historical valuation data of the asset or portfolio provided, identify trends and patterns in valuation metrics over time, and interpret the implications of these trends for investors or stakeholders.}" 
    \item Sentiment Agent: \textit{"As a sentiment equity analyst your primary responsibility is to analyze the financial news, analyst ratings and disclosures related to the underlying security; and analyze its implication and sentiment for investors or stakeholders."} 
    \item Fundamental Agent: "\textit{As a fundamental financial equity analyst your primary responsibility is to analyze the most recent 10K report provided for a company. You have access to a powerful tool that can help you extract relevant information from the 10K. Your analysis should be based solely on the information that you retrieve using this tool. You can interact with this tool using natural language queries. The tool will understand your requests and return relevant text snippets and data points from the 10K document. Keep checking if you have answered the users' question to avoid looping.}"

\end{enumerate}

\subsubsection{Agent Tools}
Each agent was equipped with specific tools and data tailored to its designated task.

For the valuation agent, we provided a computational tool to calculate the volatility and return of the stock. 

The annualized cumulative return $R_{\text{annualized}}$ is calculated (assuming 252 trading days in a year) as:
\[
R_{\text{annualized}} = \left( (1 + R_{\text{cumulative}})^{\frac{252}{n}} \right) - 1,
\]
where $R_{\text{cumulative}}$ is the cumulative return and $n$ is the number of trading days. 
Annualized volatility, $\sigma_{\text{annualized}}$ , is computed as:
\[
\sigma_{\text{annualized}} = \sigma_{\text{daily}} \times \sqrt{252},
\]
where $\sigma_{\text{daily}}$ denotes daily volatility.

For the sentiment agent, a LLM-based summarization tool was granted to condense each news item before providing an overarching sentiment analysis. This tool employs reflection-enhanced prompting, in which the model is explicitly instructed to reason, critique, and refine its summary. The guiding prompt is" "provide a concise summary along with an informed recommendation on whether to invest in this stock". The model is encouraged to reason through or evaluate the content before summarizing, often involving multiple steps like summarizing, critiquing, and refining, which leads to more coherent and insightful summaries compared to direct extractive methods. Unlike a RAG tool, a summarization tool allows the sentiment agent to learn through all the data and to provide its opinion. 

For the fundamental agent, we provided agent tool augments of both 10K/10Q data pull and RAG tool tailored for financial statement analysis for it to analyze the equity from different perspectives. 
\begin{itemize}
    \item \textbf{Fundamental Report Pull Tool:} The agent is prompted with capability to generate yfinance API calls for the specific stock he is involved. Within the tool, we include iterative API call checks to ensure the API call is correct and data is properly retrieved.  
    
    \item \textbf{Financial Report RAG Tool:} The agent can further use this tool to perform an analysis based on the fundamental reports retrieved to perform a fundamental analysis. Within the RAG tool, we leveraged context trucking based on report sections to keep relevant information and use GPT 4-o as embedding model. The tool incorporate instruction of domain expertise guide of how to analyze different section of the financial reports. The agent will use the RAG tool several times to perform an analysis answering questions relevant to cash flow and income, operations and gross margin, areas of concern and progress towards its stated objectives.     

\end{itemize}

\subsubsection{Multi-Agents Workflow}
\begin{figure} 
    \centering
    \includegraphics[width=1\linewidth]{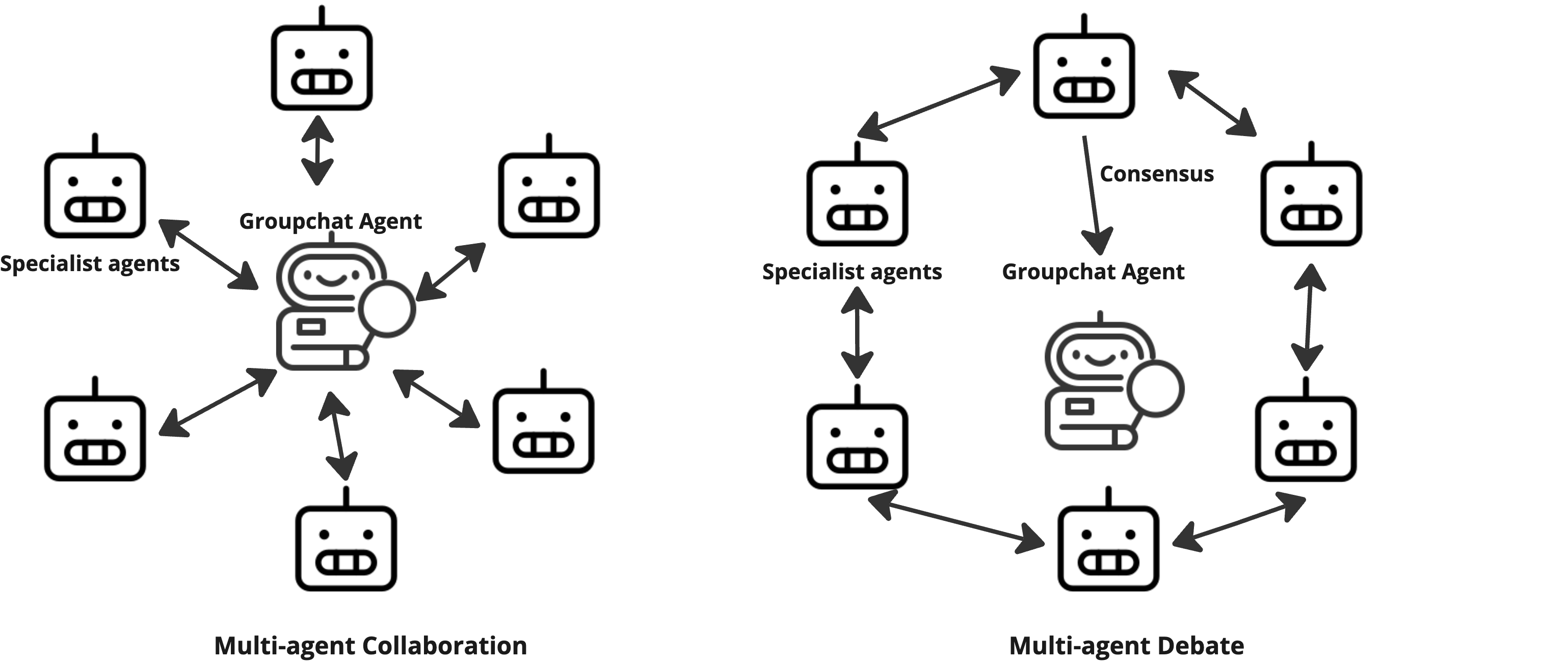}
    \caption{The AlphaAgents agentic communication workflow diagram.}
    \label{fig:alpha-agent-workflow}
\end{figure}

We utilize the Microsoft AutoGen framework \cite{wu2023autogen} as the infrastructure for our multi-agent system, leveraging its group chat and assistant agent capabilities. Autogen Studio serves as the user interface. The communication workflow is illustrated in Figure \ref{fig:alpha-agent-workflow} and the prompts for the group chat agent are provided below.

\noindent\textbf{Multi-agent Collaboration:}
To coordinate the three specialist agents, a group chat assistant was designed to consolidate their inputs and generate a collaborative stock analysis report. The group chat assistant is prompted as below: 

\noindent\textbf{Prompt:}
\textit{"You are a helpful assistant skilled at coordinating a group of other agents to solve a task. You make sure that every agent in the group chat has a chance to speak at least twice. When all agents provide their analysis, consolidate inputs of all agent into a report. 
Reply "TERMINATE" at the end when everything is done. "}

\noindent\textbf{Multi-agent Debate: }
Debate is managed via a Round Robin approach\cite{wu2023autogen}: each agent receives the query along with peer analyses, and the discussion continues until consensus is reached.

\noindent\textbf{Prompt:} 
\textit{"You are a helpful assistant skilled at coordinating a group of other agents to solve a task. You make sure that every agent in the group chat has a chance to speak at least twice. Each agent can not decide for the whole group. They are tasked with coming to a consensus. You must invoke all agents before deciding to Terminate. 
Reply "TERMINATE" at the end when everything is done." }

\subsubsection{LLM Model}
After experimenting with various GPT models, we chose GPT4o as the model for our experiments.  
\begin{figure}
    \centering
    \includegraphics[width=1\linewidth]{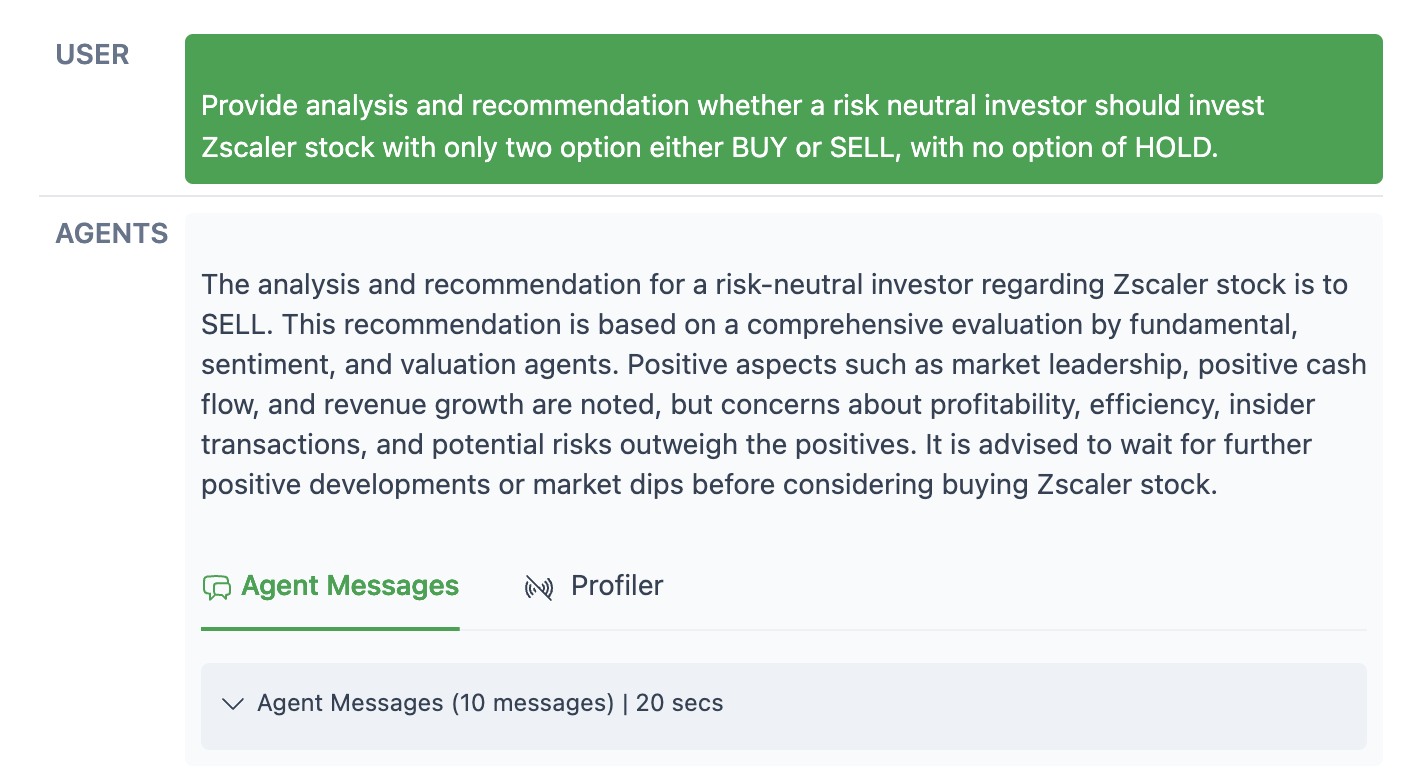}
    \caption{Multi-agent Debate Example on Zscaler}
    \label{fig:enter-label}
\end{figure}

\begin{figure}
    \centering
    \includegraphics[width=1\linewidth]{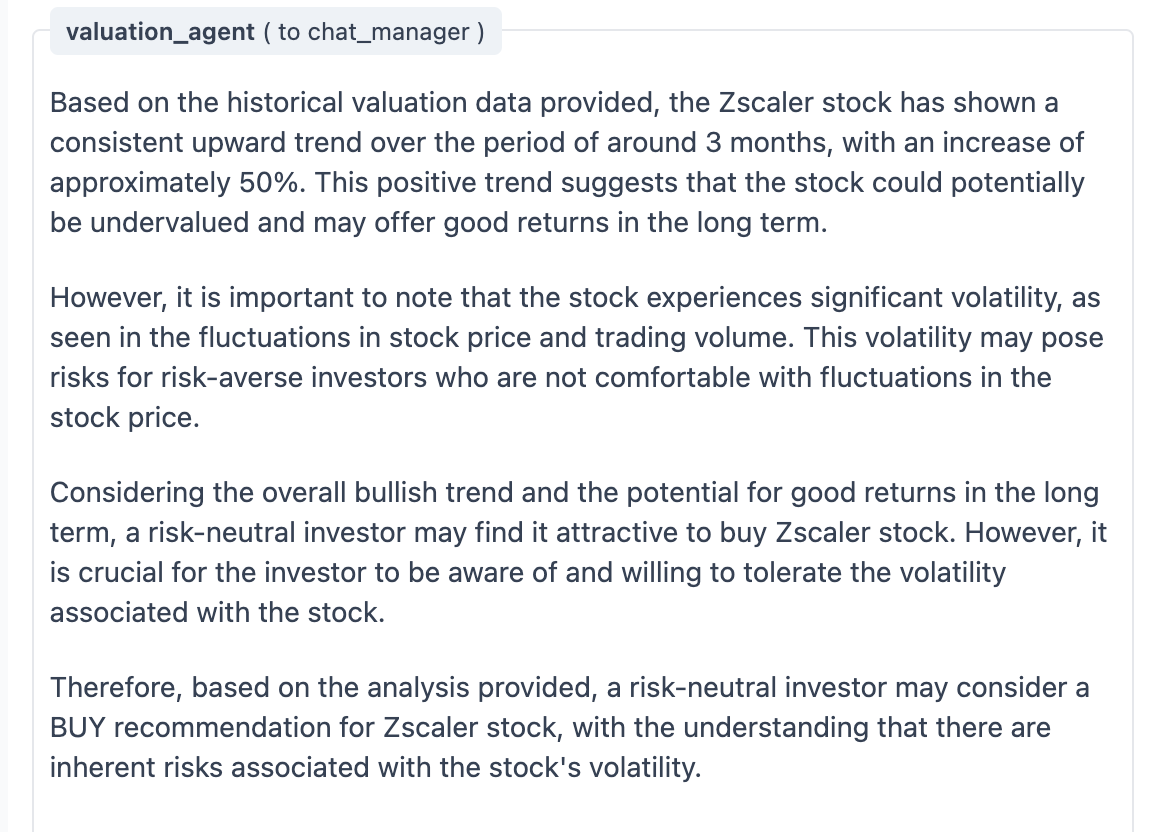}
    \includegraphics[width=1\linewidth]{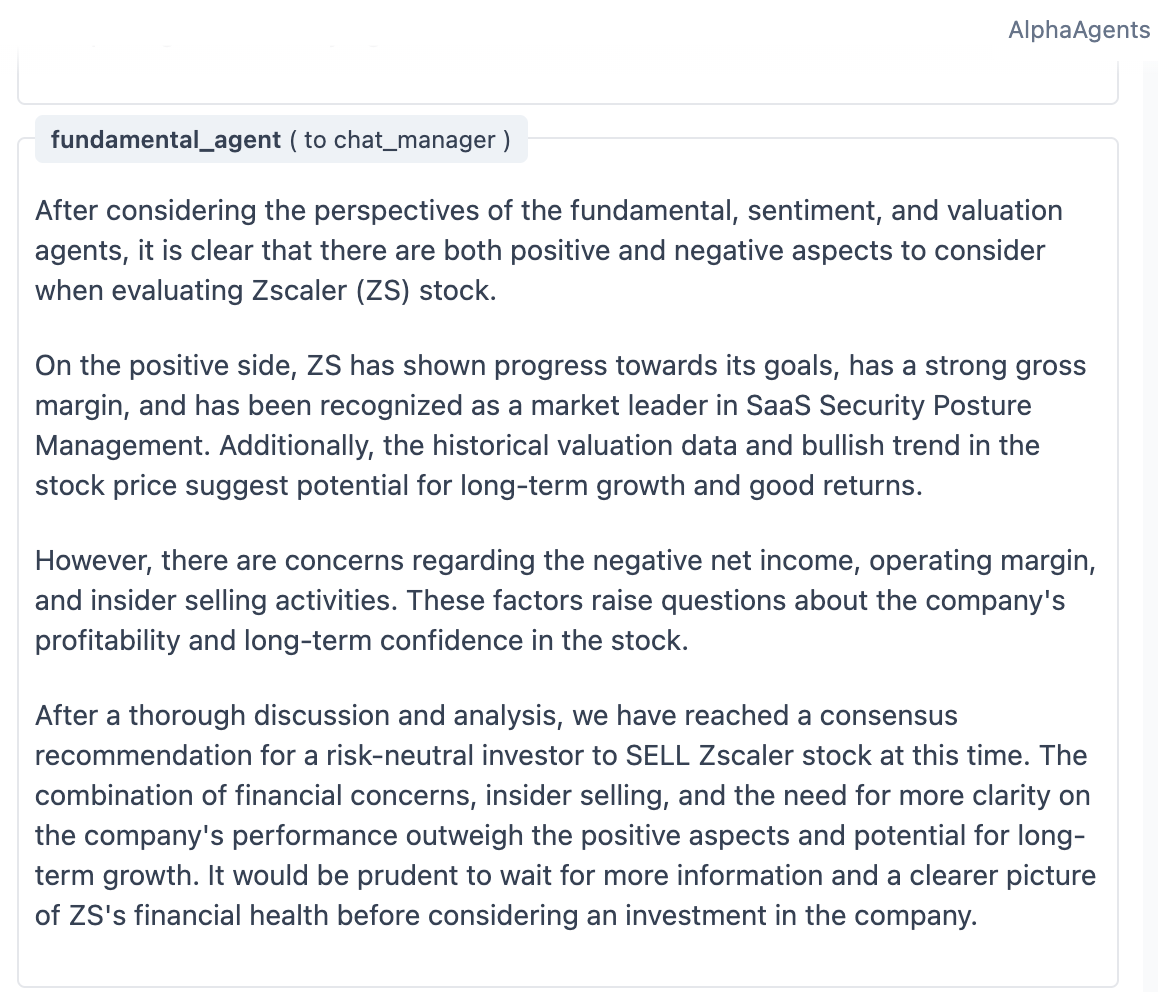}
    \caption{Multi-agent Debate Process Illustration}
    \label{fig:enter-label}
\end{figure}

\begin{figure}
    \centering
    \includegraphics[width=1\linewidth]{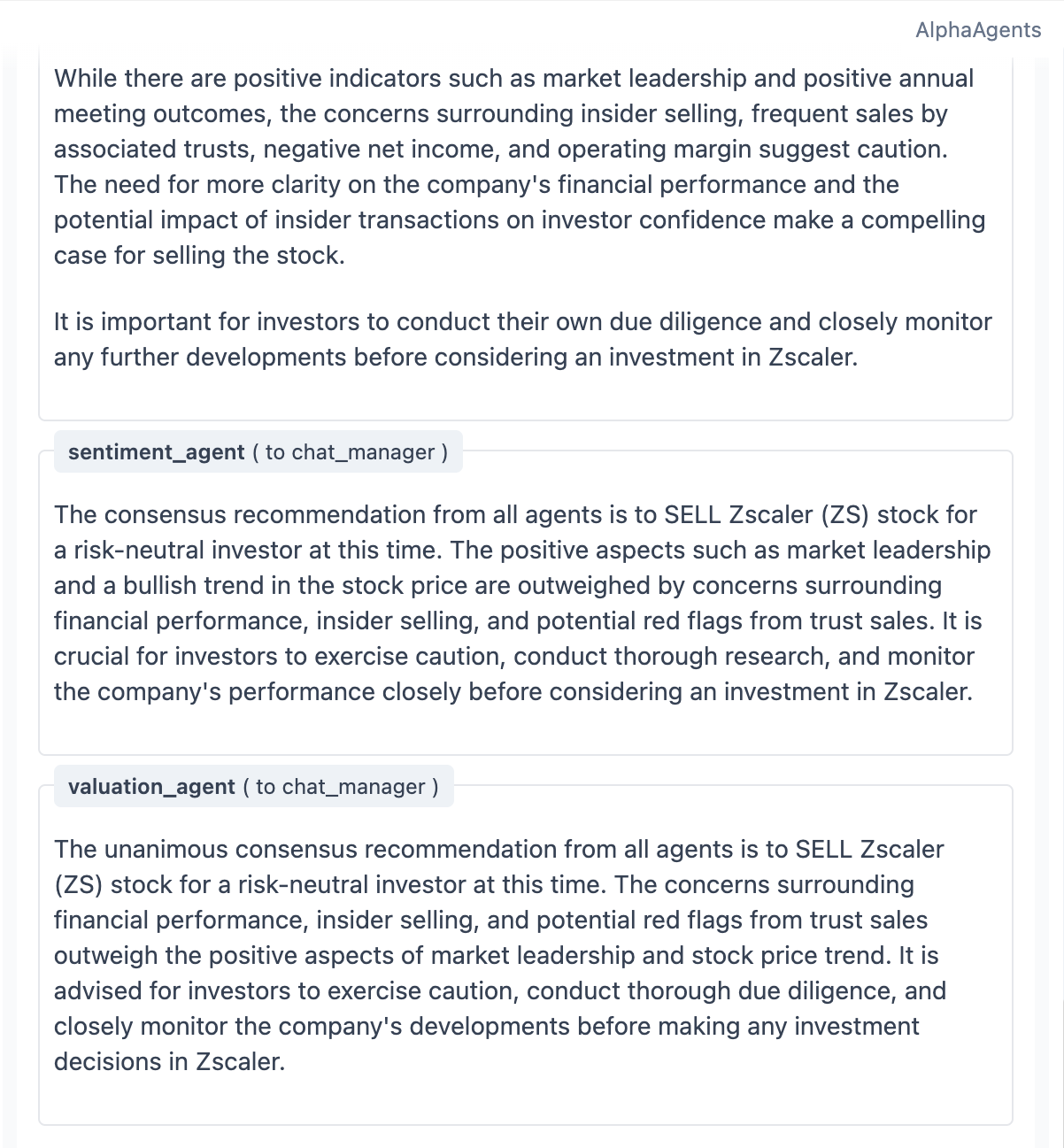}
    \caption{Multi-agent Debate Consensus Illustration}
    \label{fig:enter-label}
\end{figure}

\subsection{Impact of the Agent Risk Tolerance}
Risk tolerance plays a pivotal role in shaping investor behavior and portfolio decisions~\cite{baruah2018impact}, yet its influence on AI agent behavior remains underexplored. We illustrate this through agent responses conditioned on risk profiles: a valuation agent issues a "Sell" for Company A under a risk-averse prompt, citing high volatility despite recent outperformance, while the same agent recommends a "Buy" under a risk-neutral profile, emphasizing momentum with caution.

To model this systematically, we design agents for both risk-averse and risk-neutral scenarios using prompt engineering to embed investor traits directly into agent instructions. This mimics how human clients express preferences, enabling agents to interpret risk contextually rather than through fixed thresholds. While a risk-seeking profile was also tested, its outputs were nearly indistinguishable from the risk-neutral case and excluded from final evaluation highlighting a limitation of prompt-based differentiation for adjacent risk profiles.

\section{Evaluation Metrics}
LLM evaluation remains an evolving and challenging research area, especially to find benchmark of agent performance across diverse scenarios within a unified framework \cite{guo2024large}. There have been various automatic evaluation methods based on LLMs proposed in recent research \cite{gao2024llm}.

\subsection{RAG Evaluation}
For our multi-agent system, we deployed Arize Phoenix\footnote{\url{https://github.com/Arize-ai/phoenix}} to implement retrieval and relevance evaluation metrics for RAG \cite{gao2023retrieval} for the fundamental and sentiment agents (as they perform their tasks mainly based on RAG and summarization). Common metrics of faithfullness and relevance scores were provided as part of Phoenix pipeline. 
Evaluating the valuation agent is challenging as it performs numerical analytics with no easy ground truth. To reduce hallucination, we constructed mathematical tools for the valuation agent to calculate volatility and valuation metrics.  Also using Arize Phoenix, we constantly monitored the valuation agent and checked if it was using the mathematical tool we provided to answer the question.

\subsection{Back-testing as a Down-stream Metric}
Evaluating multi-agent debates, which involve reasoning and soundness checks, poses significant challenges. To address this, human review is employed to verify the logical coherence of the debates. For greater transparency in multi-agent reasoning, all discussion histories are logged as a supplementary output, allowing users to review and, if necessary, override the agents' conclusions.

To assess the system's performance, we conducted experiments evaluating its capabilities in both multi-agent collaboration and debating, focusing on equity analysis and stock selection as key tasks. These experiments also served as a measure of the proposed system's effectiveness in improving investment outcomes.

For this experiment, we randomly selected 15 stocks in technology sector to serve as both the stock-picking pool and the benchmark. The focus was solely on stock selection, with subsequent tasks such as portfolio diversification or optimization falling outside the scope of this study. All stocks were assigned equal weights based on agent recommendations. Future research will explore functionality to adjust stock weights according to the LLM's confidence in its recommendations, where stronger confidence (e.g., a strong BUY) would correspond to higher weight allocations.

Using the proposed multi-agent framework, we first generated a detailed stock analysis report for each stock. A debate mechanism then aggregated perspectives from the fundamental, sentiment, and valuation agents, each issuing a BUY or SELL based on their specialization and specified risk tolerance. The debate continued until consensus was reached on inclusion in the portfolio. The resulting multi-agent portfolio was constructed by modifying a benchmark portfolio. For comparison, single-agent portfolios from the fundamental and valuation agents were also created; the sentiment agent was excluded due to insufficient news coverage in some cases.

To evaluate the performance of an agent-picked portfolio, we back-tested the multi-agent performance against the benchmark portfolio and compared the risk-adjusted return and Sharpe ratio. 

The agents were trained using data and news from January 2024. Based on each agent's responses and their final decisions following debate and collaboration, we constructed the corresponding portfolios: the valuation agent portfolio, news sentiment agent portfolio, fundamental agent portfolio, and the final multi-agent portfolio, all initialized on February 1, 2024. The performance of these portfolios was monitored over a four-month period, with their risk-adjusted returns compared to evaluate effectiveness.

The risk-adjusted returns (Sharpe Ratio) is calculated using:
\[
S = \frac{R_p - R_f}{\sigma_p}
\]
Where $R_{\text{p}}$ is the expected return of portfolio, $R_{\text{f}}$ is the risk-free rate, which serves as a baseline, and $\sigma_{\text{p}}$ is the standard deviation of portfolio, which measures how much the return varys. Here for Risk free rate we are using one month treasury rate. 

\section{Result}
\subsection{Stock Analysis Report Generation}
In the experiment, the group chat agent coordinated with the specialist agents to collectively generate a stock analysis report for each stock. To provide a concrete example, below is the summary of results from the agents collaborating to produce a comprehensive analysis report for a sample company Z.
\begin{enumerate}
\item\textbf{Positive Indicators}
Overall Performance: Company Z's stock increased by approximately 13.56\% from January 2nd to
January 30th, outperforming the SP 500's increase of roughly 3.85\%. The stock shows higher
volatility with notable price swings, which could lead to higher returns if the market moves favorably.

Market Leadership: Company Z has been recognized as the exclusive leader in the Forrester Wave for
SaaS Security Posture Management for Q4 2023, highlighting its strong market position and
commitment to providing advanced security solutions.

Annual Meeting Outcomes: The election of Jay Chaudhry as a Class III director and the ratification
of PricewaterhouseCoopers LLP as the independent auditor suggest stability and continuity in
leadership and financial oversight. The approval of executive compensation by stockholders
indicates confidence in the management team.

Progress Towards Goals: Company Z has made significant progress towards its 2023 goals with
initiatives such as the Senior Leadership Program, partnerships with executive coaching
organizations, and offering tuition reimbursement for employee career growth. The company also
experienced revenue growth, expanded its customer base, and streamlined operations through a
restructuring plan.

\item\textbf{Concerns}
Higher Risk: The higher volatility of Company Z's stock entails greater risk, with the price equally likely to
go down as well as up. The trade volumes for Company Z's stock are relatively lower than the SP 500,
potentially leading to less liquidity and difficulty in buying or selling the stock.
Insider Selling: Recent sales by multiple directors, including Amit Sinha and Karen Blasing, might
indicate that insiders believe the stock is fairly valued or overvalued. The proposed sales of Company Z
shares by the Sinha Family Incentive Trust and the Sinha Education Excellence Trust raise
concerns about the long-term confidence in the company's stock.
Financial Metrics: Company Z's cash flow for the year 2023 is significantly higher than its net income, as
reported in the SEC Filing 10-K. The operating margin for Company Z is -14.5\%, indicating that the
company is operating at a loss. Despite a net loss in the fiscal year 2023, Company Z has shown an
improvement in revenue growth, with the net loss amount decreasing compared to the previous
fiscal year.

\item\textbf{Conclusion}
Investment Considerations: Investors with a higher risk tolerance may view Company Z's stock as a good entry point, given its potential for higher returns balanced against the risk. Conversely, more conservative investors may prefer to consider other alternative opportunities or wait for greater price stability. While company Z's market leadership and positive outcomes from annual meeting are encouraging, recent insider selling and frequent transactions by associated trusts suggest caution. Close monitoring the stock is recommended, with
consideration given to purchasing during market dips or following further positive developments.
\end{enumerate}

\subsection{Stock Selection Result}
\subsubsection{Risk-Seeking/Risk-Neutral Agents:}
The \textbf{valuation agent} uses the benchmark portfolio directly, implying it does not differentiate its selection from the market baseline. The \textbf{fundamental agent} expands upon the benchmark by including a broader set of stocks, suggesting a strategy that emphasizes diversification and potentially considers additional financial fundamentals or growth signals beyond the baseline. The \textbf{multi-agent system} includes most of the fundamental agent’s picks but appears slightly more curated, indicating coordination among agents may help refine or consolidate decisions while retaining broader exposure.

\subsubsection{Risk-Averse Agents:}
The \textbf{valuation agent} adopts a strategy that narrows the selection, excluding several stocks present in the risk-neutral set. This suggests a more conservative filtering based on valuation criteria, likely emphasizing lower volatility or more stable fundamentals. The \textbf{fundamental agent} focuses on a small subset of stocks, favoring those that may offer stability or strong balance sheets, consistent with a defensive investment posture. The \textbf{multi-agent strategy} further consolidates the portfolio to a handful of overlapping picks, reflecting cautious consensus among valuation and fundamental signals in high-risk scenarios.

\begin{figure}
    \centering
    \includegraphics[width=0.7\linewidth]{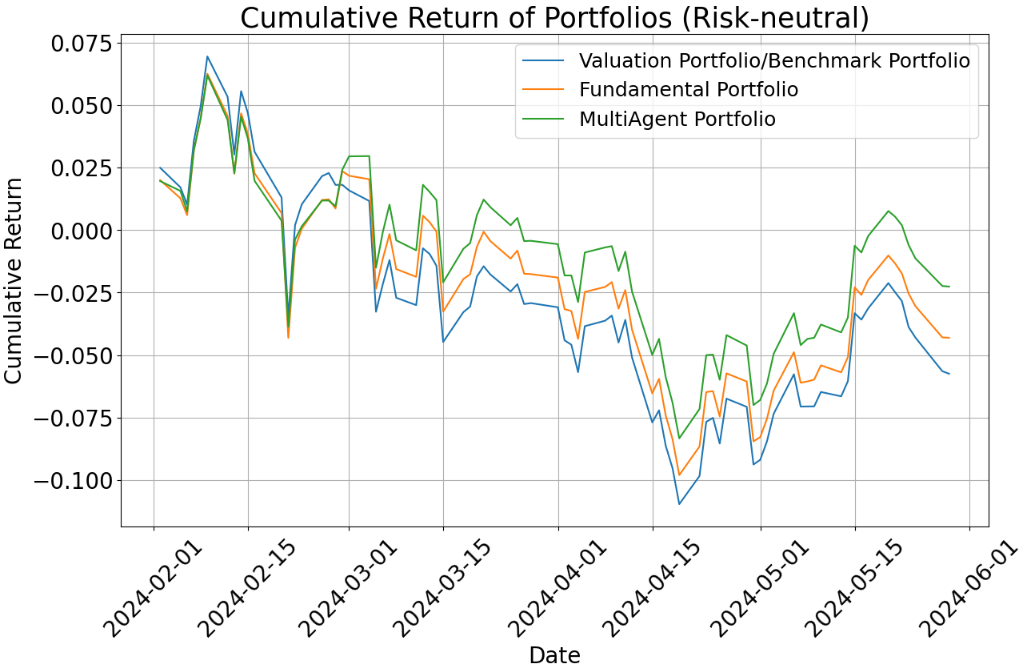}
    \includegraphics[width=0.7\linewidth]{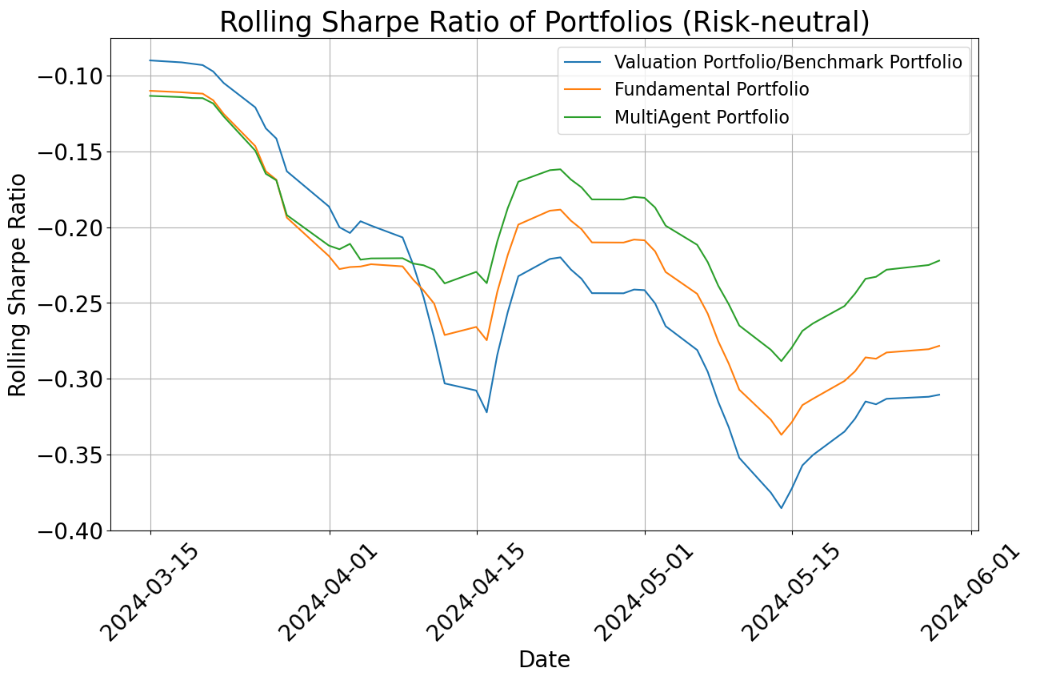}
    \caption{Risk Neutral Portfolio Performance: the multi-agent portfolio outperforms both single-agent portfolios in terms of cumulative return and rolling Sharpe ratio}
    \label{fig:risk-neutral-sharpe}
    \label{fig:risk-neutral-sharpe}
\end{figure}

\begin{figure}
    \centering
    \includegraphics[width=0.7\linewidth]{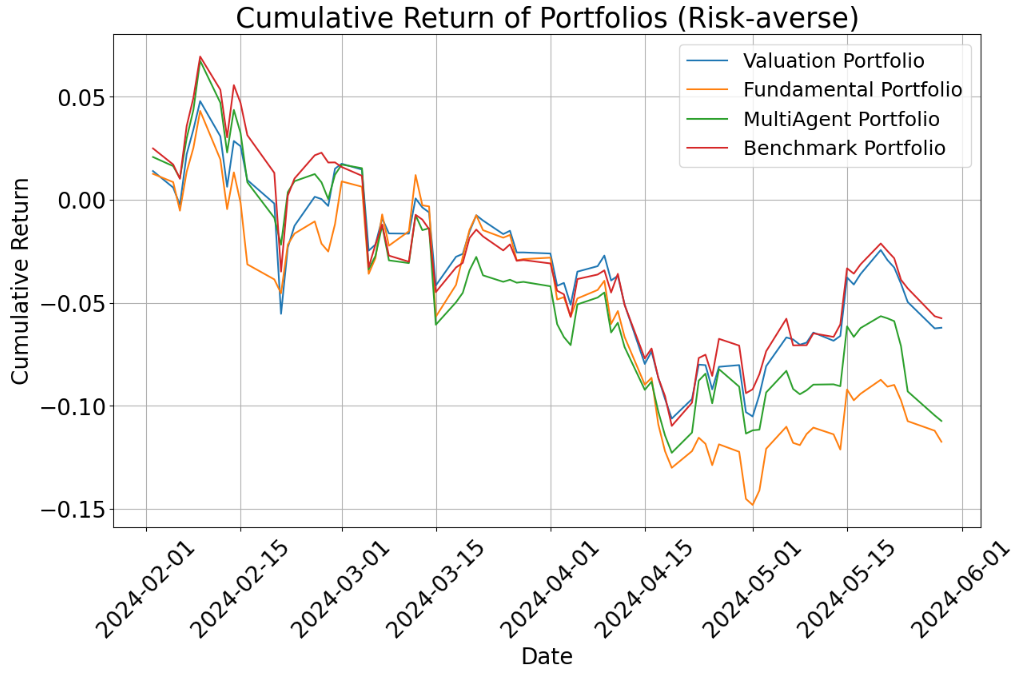}
    \includegraphics[width=0.7\linewidth]{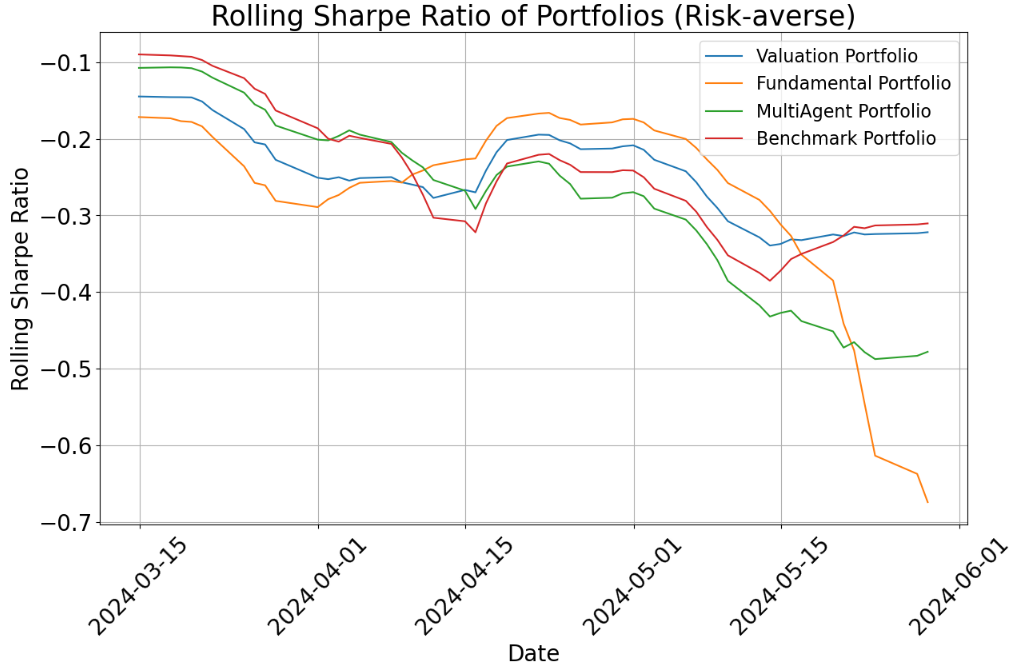}
    \caption{Risk Averse Portfolio Performance: all portfolios experience larger drawdowns}
    \label{fig:risk-averse-sharpe}
\end{figure}

\begin{figure}
    \centering
    \includegraphics[width=0.7\linewidth]{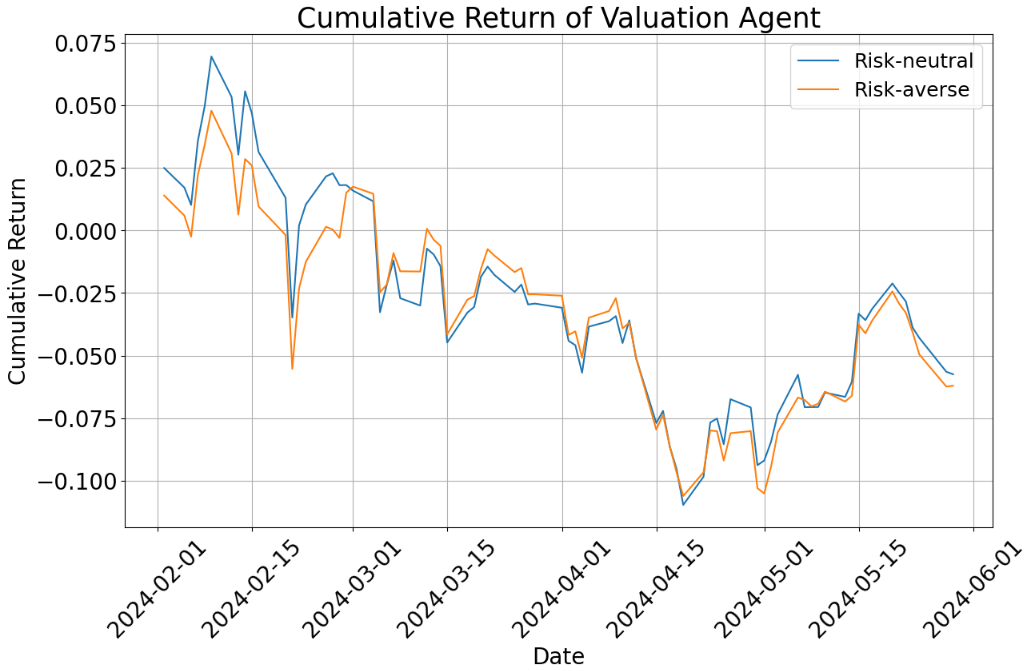}
    \includegraphics[width=0.7\linewidth]{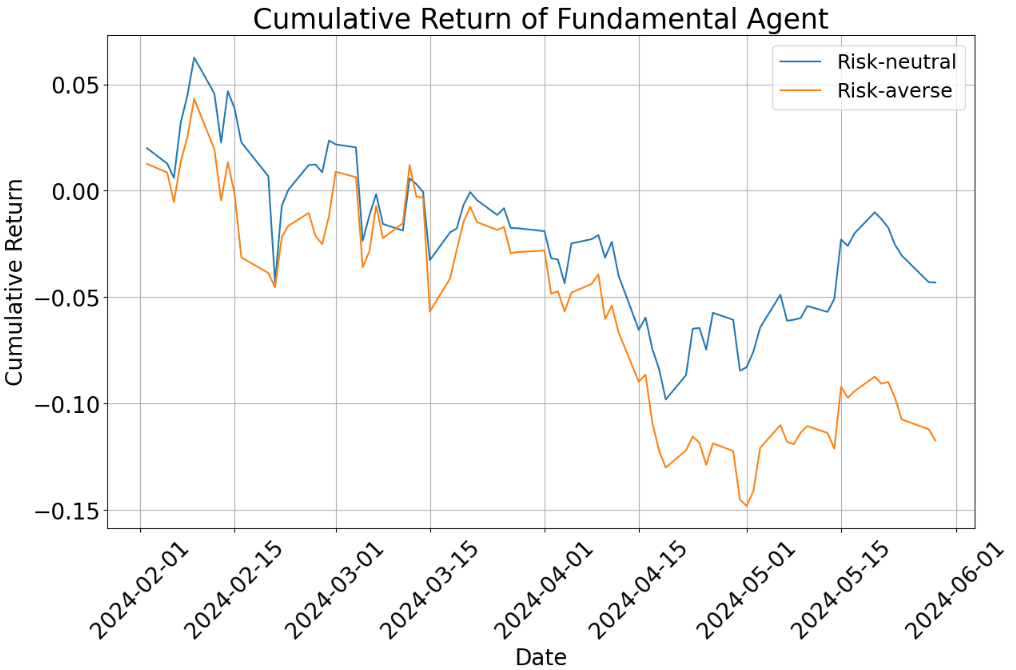}
    \includegraphics[width=0.7\linewidth]{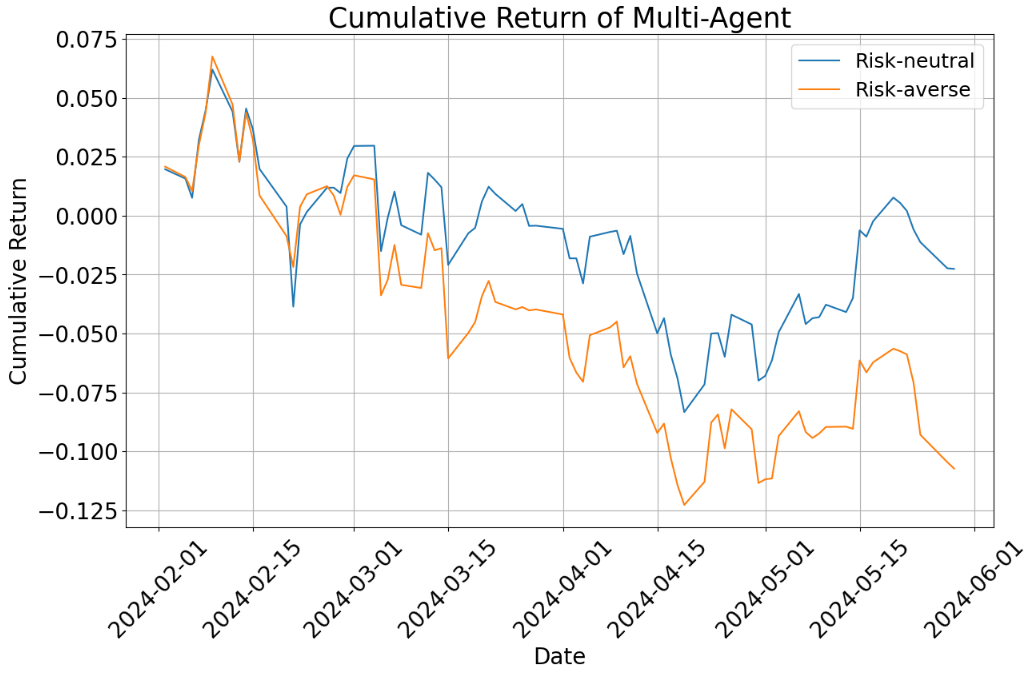}
    \caption{Risk neutral vs Risk averse: Comparing risk-neutral and risk-averse strategies, risk-neutral portfolios consistently achieve higher cumulative returns across all agent types}
    \label{fig:risk-neutral-vs-averse}
\end{figure}

\subsection{Back-Test Results}
The portfolios constructed by a valuation agent and a fundamental agent are compared to those selected through a multi-agent framework. Portfolio selection occurred on February 1, 2024, with performance measured over the subsequent four months. To facilitate a dynamic performance assessment, a rolling Sharpe ratio was calculated throughout the evaluation period. 

The rolling Sharpe Ratio is defined as:
\[S_{\text{rolling}}(t) = \frac{\overline{R_{p, t-w+1:t}} - R_f}{\sigma_{p, t-w+1:t}}\]
where the mean portfolio return over the window $w$ is:
\[\overline{R_{p, t-w+1:t}} = \frac{1}{w} \sum_{i=t-w+1}^{t} R_{p, i}\]
and the corresponding standard deviation is:
\[\sigma_{p, t-w+1:t} = \sqrt{\frac{1}{w-1} \sum_{i=t-w+1}^{t} (R_{p, i} - \overline{R_{p, t-w+1:t}})^2}\]
Results are presented in Figures \ref{fig:risk-neutral-sharpe}, \ref{fig:risk-averse-sharpe} and \ref{fig:risk-neutral-vs-averse}. In the risk-neutral setting, the proposed multi-agent outperforms the benchmark as well as the single agent framework during the testing window. The sentiment and valuation agents, which analyze on a short historical horizon of 1 to 3 months, are more effective for short-term forecasting. In contrast, the fundamental agent focuses on the long-term potential by analyzing company fundamentals, such as information from 10-K filings. By combining these perspectives, the multi-agent system effectively balances short-term and long-term insights. As a result, it achieves superior performance compared to the single agent framework and outperforms all other benchmark portfolios. 

In the risk-averse setting, all agent-selected portfolios followed a conservative investment strategy, leading to underperformance relative to the benchmark. This outcome was partly due to the strong gains of the technology sector during the testing period. Given that technology stocks are typically more volatile, they were largely excluded from the risk-averse portfolios. While this reduced exposure effectively limited downside risk, it also constrained the potential for upside gains. This disparity highlights the classic trade-off between risk mitigation and return potential: during a predominantly bullish market, portfolios with higher risk tolerance tend to outperform, particularly those including high-growth sectors such as technology. In contrast, conservative portfolios prioritize lower volatility, which inherently limits their ability to capitalize on robust market rallies.

However, as shown in Figure \ref{fig:risk-averse-sharpe}, the multi-agent portfolio achieved relatively stronger performance than both the valuation and fundamental portfolios, particularly during the early stages of the testing period. This suggests that the integration of conservative strategies from both the valuation and fundamental agents allowed the multi-agent framework to more effectively balance risk and return. While all three approaches maintained a risk-averse posture, the multi-agent portfolio exhibited slightly lower volatility and reduced drawdowns, indicating an ability to mitigate losses more effectively than the individual strategies. Although the analysis is limited in scope, the observed reduction in downside risk support the framework’s intended conservative design.

Figure \ref{fig:risk-neutral-vs-averse} shows that among all the scenarios the risk averse agent will lead to lower returns as it cautiously selects more stable stock and yields to lower returns under a bullish market, which is an obvious artifact of the risk aversion. On the other hand, the results exhibit that the multi-agent portfolio shows an ability to fill a gap between the fundamental agent and the valuation agent, and balance the risk tolerance impact between these agents as well.  

\section{Conclusion}
We presented AlphaAgents, a modular multi-agent debate framework for stock analysis and selection, demonstrating that coordinated agent reasoning can enhance decision quality and mitigate human cognitive biases. Comprising specialized fundamental, sentiment, and valuation agents, the system synthesizes diverse perspectives through structured collaboration, yielding robust stock assessments and explainable outputs. While not yet a full portfolio optimization engine, AlphaAgents represents a foundational step toward agentic investment systems.

Our backtesting results indicate that multi-agent debate improves analytical rigor, especially in scenarios involving conflicting signals or partial agent reliability. Moreover, agents' discussion logs provide transparent reasoning trails, addressing a key challenge in evaluating collaborative AI systems.

AlphaAgents aligns with established portfolio construction frameworks and discretionary investment workflows. While currently focused on stock selection, it can serve as a modular input to models like Mean-Variance Optimization or Black-Litterman, by supplying agent-driven signals for return estimation and scenario analysis. The system’s structure mirrors the way discretionary managers synthesize diverse research inputs.

The debate mechanism also reflects investment committee-style reasoning, where differing perspectives are reconciled through discussion. This makes AlphaAgents well-suited for augmenting or automating human-in-the-loop decision-making in institutional settings. With improved infrastructure and better data integration, such frameworks can become scalable, transparent components of future portfolio management systems.

\section{Acknowledgement}
The views expressed here are those of the authors alone and not of BlackRock, Inc.


\bibliographystyle{ACM-Reference-Format}
\bibliography{sample-base}


\begin{thebibliography}{34}


\ifx \showCODEN    \undefined \def \showCODEN     #1{\unskip}     \fi
\ifx \showDOI      \undefined \def \showDOI       #1{#1}\fi
\ifx \showISBNx    \undefined \def \showISBNx     #1{\unskip}     \fi
\ifx \showISBNxiii \undefined \def \showISBNxiii  #1{\unskip}     \fi
\ifx \showISSN     \undefined \def \showISSN      #1{\unskip}     \fi
\ifx \showLCCN     \undefined \def \showLCCN      #1{\unskip}     \fi
\ifx \shownote     \undefined \def \shownote      #1{#1}          \fi
\ifx \showarticletitle \undefined \def \showarticletitle #1{#1}   \fi
\ifx \showURL      \undefined \def \showURL       {\relax}        \fi
\providecommand\bibfield[2]{#2}
\providecommand\bibinfo[2]{#2}
\providecommand\natexlab[1]{#1}
\providecommand\showeprint[2][]{arXiv:#2}

\bibitem[Agashe et~al\mbox{.}(2023)]%
        {agashe2023evaluating}
\bibfield{author}{\bibinfo{person}{Saaket Agashe}, \bibinfo{person}{Yue Fan}, {and} \bibinfo{person}{Xin~Eric Wang}.} \bibinfo{year}{2023}\natexlab{}.
\newblock \showarticletitle{Evaluating multi-agent coordination abilities in large language models}.
\newblock \bibinfo{journal}{\emph{arXiv preprint arXiv:2310.03903}} (\bibinfo{year}{2023}).
\newblock


\bibitem[An et~al\mbox{.}(2024)]%
        {an2024finverse}
\bibfield{author}{\bibinfo{person}{Siyu An}, \bibinfo{person}{Qin Li}, \bibinfo{person}{Junru Lu}, \bibinfo{person}{Di Yin}, {and} \bibinfo{person}{Xing Sun}.} \bibinfo{year}{2024}\natexlab{}.
\newblock \showarticletitle{FinVerse: An Autonomous Agent System for Versatile Financial Analysis}.
\newblock \bibinfo{journal}{\emph{arXiv preprint arXiv:2406.06379}} (\bibinfo{year}{2024}).
\newblock


\bibitem[Antony(2020)]%
        {antony2020behavioral}
\bibfield{author}{\bibinfo{person}{Anu Antony}.} \bibinfo{year}{2020}\natexlab{}.
\newblock \showarticletitle{Behavioral finance and portfolio management: Review of theory and literature}.
\newblock \bibinfo{journal}{\emph{Journal of Public Affairs}} \bibinfo{volume}{20}, \bibinfo{number}{2} (\bibinfo{year}{2020}), \bibinfo{pages}{e1996}.
\newblock


\bibitem[Baker and Nofsinger(2010)]%
        {baker2010behavioral}
\bibfield{author}{\bibinfo{person}{H~Kent Baker} {and} \bibinfo{person}{John~R Nofsinger}.} \bibinfo{year}{2010}\natexlab{}.
\newblock \bibinfo{booktitle}{\emph{Behavioral finance: investors, corporations, and markets}}. Vol.~\bibinfo{volume}{6}.
\newblock \bibinfo{publisher}{John Wiley \& Sons}.
\newblock


\bibitem[Baruah and Parikh(2018)]%
        {baruah2018impact}
\bibfield{author}{\bibinfo{person}{Mitali Baruah} {and} \bibinfo{person}{Abhishek~Kiritkumar Parikh}.} \bibinfo{year}{2018}\natexlab{}.
\newblock \showarticletitle{Impact of risk tolerance and demographic factors on financial investment decision}.
\newblock \bibinfo{journal}{\emph{International Journal of Financial Management}} \bibinfo{volume}{8}, \bibinfo{number}{1} (\bibinfo{year}{2018}), \bibinfo{pages}{36--48}.
\newblock


\bibitem[Caparrelli et~al\mbox{.}(2004)]%
        {caparrelli2004herding}
\bibfield{author}{\bibinfo{person}{Franco Caparrelli}, \bibinfo{person}{Anna~Maria D'Arcangelis}, {and} \bibinfo{person}{Alexander Cassuto}.} \bibinfo{year}{2004}\natexlab{}.
\newblock \showarticletitle{Herding in the Italian stock market: a case of behavioral finance}.
\newblock \bibinfo{journal}{\emph{The Journal of Behavioral Finance}} \bibinfo{volume}{5}, \bibinfo{number}{4} (\bibinfo{year}{2004}), \bibinfo{pages}{222--230}.
\newblock


\bibitem[Davies and Brooks(2017)]%
        {davies2017practical}
\bibfield{author}{\bibinfo{person}{Greg~B Davies} {and} \bibinfo{person}{Peter Brooks}.} \bibinfo{year}{2017}\natexlab{}.
\newblock \showarticletitle{Practical Challenges of Implementing Behavioral Finance: Reflections from the Field}.
\newblock \bibinfo{journal}{\emph{Financial Behavior: Players, Services, Products, and Markets}} (\bibinfo{year}{2017}), \bibinfo{pages}{542--560}.
\newblock


\bibitem[De~Bondt and Thaler(1995)]%
        {de1995financial}
\bibfield{author}{\bibinfo{person}{Werner~FM De~Bondt} {and} \bibinfo{person}{Richard~H Thaler}.} \bibinfo{year}{1995}\natexlab{}.
\newblock \showarticletitle{Financial decision-making in markets and firms: A behavioral perspective}.
\newblock \bibinfo{journal}{\emph{Handbooks in operations research and management science}}  \bibinfo{volume}{9} (\bibinfo{year}{1995}), \bibinfo{pages}{385--410}.
\newblock


\bibitem[Dong et~al\mbox{.}(2022)]%
        {dong2022survey}
\bibfield{author}{\bibinfo{person}{Qingxiu Dong}, \bibinfo{person}{Lei Li}, \bibinfo{person}{Damai Dai}, \bibinfo{person}{Ce Zheng}, \bibinfo{person}{Zhiyong Wu}, \bibinfo{person}{Baobao Chang}, \bibinfo{person}{Xu Sun}, \bibinfo{person}{Jingjing Xu}, {and} \bibinfo{person}{Zhifang Sui}.} \bibinfo{year}{2022}\natexlab{}.
\newblock \showarticletitle{A survey on in-context learning}.
\newblock \bibinfo{journal}{\emph{arXiv preprint arXiv:2301.00234}} (\bibinfo{year}{2022}).
\newblock


\bibitem[Du et~al\mbox{.}(2023)]%
        {du2023improving}
\bibfield{author}{\bibinfo{person}{Yilun Du}, \bibinfo{person}{Shuang Li}, \bibinfo{person}{Antonio Torralba}, \bibinfo{person}{Joshua~B Tenenbaum}, {and} \bibinfo{person}{Igor Mordatch}.} \bibinfo{year}{2023}\natexlab{}.
\newblock \showarticletitle{Improving factuality and reasoning in language models through multiagent debate}.
\newblock \bibinfo{journal}{\emph{arXiv preprint arXiv:2305.14325}} (\bibinfo{year}{2023}).
\newblock


\bibitem[Evensky(2017)]%
        {evensky2017applications}
\bibfield{author}{\bibinfo{person}{Harold Evensky}.} \bibinfo{year}{2017}\natexlab{}.
\newblock \showarticletitle{Applications of Client Behavior: A Practitioner's Perspective}.
\newblock \bibinfo{journal}{\emph{Financial Behavior: Players, Services, Products, and Markets. H. Kent Baker, Greg Filbeck, and Victor Ricciardi, editors}} (\bibinfo{year}{2017}), \bibinfo{pages}{523--541}.
\newblock


\bibitem[Fatouros et~al\mbox{.}(2025)]%
        {fatouros2025marketsenseai}
\bibfield{author}{\bibinfo{person}{George Fatouros}, \bibinfo{person}{Kostas Metaxas}, \bibinfo{person}{John Soldatos}, {and} \bibinfo{person}{Manos Karathanassis}.} \bibinfo{year}{2025}\natexlab{}.
\newblock \showarticletitle{Marketsenseai 2.0: Enhancing stock analysis through llm agents}.
\newblock \bibinfo{journal}{\emph{arXiv preprint arXiv:2502.00415}} (\bibinfo{year}{2025}).
\newblock


\bibitem[Fromlet(2001)]%
        {fromlet2001behavioral}
\bibfield{author}{\bibinfo{person}{Hubert Fromlet}.} \bibinfo{year}{2001}\natexlab{}.
\newblock \showarticletitle{Behavioral finance-theory and practical application: Systematic analysis of departures from the homo oeconomicus paradigm are essential for realistic financial research and analysis}.
\newblock \bibinfo{journal}{\emph{Business economics}} (\bibinfo{year}{2001}), \bibinfo{pages}{63--69}.
\newblock


\bibitem[Gao et~al\mbox{.}(2024)]%
        {gao2024llm}
\bibfield{author}{\bibinfo{person}{Mingqi Gao}, \bibinfo{person}{Xinyu Hu}, \bibinfo{person}{Jie Ruan}, \bibinfo{person}{Xiao Pu}, {and} \bibinfo{person}{Xiaojun Wan}.} \bibinfo{year}{2024}\natexlab{}.
\newblock \showarticletitle{Llm-based nlg evaluation: Current status and challenges}.
\newblock \bibinfo{journal}{\emph{arXiv preprint arXiv:2402.01383}} (\bibinfo{year}{2024}).
\newblock


\bibitem[Gao et~al\mbox{.}(2023)]%
        {gao2023retrieval}
\bibfield{author}{\bibinfo{person}{Yunfan Gao}, \bibinfo{person}{Yun Xiong}, \bibinfo{person}{Xinyu Gao}, \bibinfo{person}{Kangxiang Jia}, \bibinfo{person}{Jinliu Pan}, \bibinfo{person}{Yuxi Bi}, \bibinfo{person}{Yi Dai}, \bibinfo{person}{Jiawei Sun}, {and} \bibinfo{person}{Haofen Wang}.} \bibinfo{year}{2023}\natexlab{}.
\newblock \showarticletitle{Retrieval-augmented generation for large language models: A survey}.
\newblock \bibinfo{journal}{\emph{arXiv preprint arXiv:2312.10997}} (\bibinfo{year}{2023}).
\newblock


\bibitem[Guo et~al\mbox{.}(2024)]%
        {guo2024large}
\bibfield{author}{\bibinfo{person}{Taicheng Guo}, \bibinfo{person}{Xiuying Chen}, \bibinfo{person}{Yaqi Wang}, \bibinfo{person}{Ruidi Chang}, \bibinfo{person}{Shichao Pei}, \bibinfo{person}{Nitesh~V Chawla}, \bibinfo{person}{Olaf Wiest}, {and} \bibinfo{person}{Xiangliang Zhang}.} \bibinfo{year}{2024}\natexlab{}.
\newblock \showarticletitle{Large language model based multi-agents: A survey of progress and challenges}.
\newblock \bibinfo{journal}{\emph{arXiv preprint arXiv:2402.01680}} (\bibinfo{year}{2024}).
\newblock


\bibitem[Kahneman and Tversky(2013)]%
        {kahneman2013prospect}
\bibfield{author}{\bibinfo{person}{Daniel Kahneman} {and} \bibinfo{person}{Amos Tversky}.} \bibinfo{year}{2013}\natexlab{}.
\newblock \showarticletitle{Prospect theory: An analysis of decision under risk}.
\newblock In \bibinfo{booktitle}{\emph{Handbook of the fundamentals of financial decision making: Part I}}. \bibinfo{publisher}{World Scientific}, \bibinfo{pages}{99--127}.
\newblock


\bibitem[Lee et~al\mbox{.}(2020)]%
        {lee2020maps}
\bibfield{author}{\bibinfo{person}{Jinho Lee}, \bibinfo{person}{Raehyun Kim}, \bibinfo{person}{Seok-Won Yi}, {and} \bibinfo{person}{Jaewoo Kang}.} \bibinfo{year}{2020}\natexlab{}.
\newblock \showarticletitle{MAPS: Multi-agent reinforcement learning-based portfolio management system}.
\newblock \bibinfo{journal}{\emph{arXiv preprint arXiv:2007.05402}} (\bibinfo{year}{2020}).
\newblock


\bibitem[Li et~al\mbox{.}(2024)]%
        {li2024more}
\bibfield{author}{\bibinfo{person}{Junyou Li}, \bibinfo{person}{Qin Zhang}, \bibinfo{person}{Yangbin Yu}, \bibinfo{person}{Qiang Fu}, {and} \bibinfo{person}{Deheng Ye}.} \bibinfo{year}{2024}\natexlab{}.
\newblock \showarticletitle{More agents is all you need}.
\newblock \bibinfo{journal}{\emph{arXiv preprint arXiv:2402.05120}} (\bibinfo{year}{2024}).
\newblock


\bibitem[Li et~al\mbox{.}(2023)]%
        {li2023api}
\bibfield{author}{\bibinfo{person}{Minghao Li}, \bibinfo{person}{Yingxiu Zhao}, \bibinfo{person}{Bowen Yu}, \bibinfo{person}{Feifan Song}, \bibinfo{person}{Hangyu Li}, \bibinfo{person}{Haiyang Yu}, \bibinfo{person}{Zhoujun Li}, \bibinfo{person}{Fei Huang}, {and} \bibinfo{person}{Yongbin Li}.} \bibinfo{year}{2023}\natexlab{}.
\newblock \showarticletitle{Api-bank: A comprehensive benchmark for tool-augmented llms}.
\newblock \bibinfo{journal}{\emph{arXiv preprint arXiv:2304.08244}} (\bibinfo{year}{2023}).
\newblock


\bibitem[Ma et~al\mbox{.}(2023)]%
        {ma2023multi}
\bibfield{author}{\bibinfo{person}{Cong Ma}, \bibinfo{person}{Jiangshe Zhang}, \bibinfo{person}{Zongxin Li}, {and} \bibinfo{person}{Shuang Xu}.} \bibinfo{year}{2023}\natexlab{}.
\newblock \showarticletitle{Multi-agent deep reinforcement learning algorithm with trend consistency regularization for portfolio management}.
\newblock \bibinfo{journal}{\emph{Neural Computing and Applications}} \bibinfo{volume}{35}, \bibinfo{number}{9} (\bibinfo{year}{2023}), \bibinfo{pages}{6589--6601}.
\newblock


\bibitem[Naveed et~al\mbox{.}(2023)]%
        {naveed2023comprehensive}
\bibfield{author}{\bibinfo{person}{Humza Naveed}, \bibinfo{person}{Asad~Ullah Khan}, \bibinfo{person}{Shi Qiu}, \bibinfo{person}{Muhammad Saqib}, \bibinfo{person}{Saeed Anwar}, \bibinfo{person}{Muhammad Usman}, \bibinfo{person}{Naveed Akhtar}, \bibinfo{person}{Nick Barnes}, {and} \bibinfo{person}{Ajmal Mian}.} \bibinfo{year}{2023}\natexlab{}.
\newblock \showarticletitle{A comprehensive overview of large language models}.
\newblock \bibinfo{journal}{\emph{arXiv preprint arXiv:2307.06435}} (\bibinfo{year}{2023}).
\newblock


\bibitem[Nie et~al\mbox{.}(2024)]%
        {nie2024survey}
\bibfield{author}{\bibinfo{person}{Yuqi Nie}, \bibinfo{person}{Yaxuan Kong}, \bibinfo{person}{Xiaowen Dong}, \bibinfo{person}{John~M Mulvey}, \bibinfo{person}{H~Vincent Poor}, \bibinfo{person}{Qingsong Wen}, {and} \bibinfo{person}{Stefan Zohren}.} \bibinfo{year}{2024}\natexlab{}.
\newblock \showarticletitle{A Survey of Large Language Models for Financial Applications: Progress, Prospects and Challenges}.
\newblock \bibinfo{journal}{\emph{arXiv preprint arXiv:2406.11903}} (\bibinfo{year}{2024}).
\newblock


\bibitem[Renze and Guven(2024)]%
        {renze2024self}
\bibfield{author}{\bibinfo{person}{Matthew Renze} {and} \bibinfo{person}{Erhan Guven}.} \bibinfo{year}{2024}\natexlab{}.
\newblock \showarticletitle{Self-Reflection in LLM Agents: Effects on Problem-Solving Performance}.
\newblock \bibinfo{journal}{\emph{arXiv preprint arXiv:2405.06682}} (\bibinfo{year}{2024}).
\newblock


\bibitem[Sahoo et~al\mbox{.}(2024)]%
        {sahoo2024systematic}
\bibfield{author}{\bibinfo{person}{Pranab Sahoo}, \bibinfo{person}{Ayush~Kumar Singh}, \bibinfo{person}{Sriparna Saha}, \bibinfo{person}{Vinija Jain}, \bibinfo{person}{Samrat Mondal}, {and} \bibinfo{person}{Aman Chadha}.} \bibinfo{year}{2024}\natexlab{}.
\newblock \showarticletitle{A systematic survey of prompt engineering in large language models: Techniques and applications}.
\newblock \bibinfo{journal}{\emph{arXiv preprint arXiv:2402.07927}} (\bibinfo{year}{2024}).
\newblock


\bibitem[Song et~al\mbox{.}(2023)]%
        {song2023llm}
\bibfield{author}{\bibinfo{person}{Chan~Hee Song}, \bibinfo{person}{Jiaman Wu}, \bibinfo{person}{Clayton Washington}, \bibinfo{person}{Brian~M Sadler}, \bibinfo{person}{Wei-Lun Chao}, {and} \bibinfo{person}{Yu Su}.} \bibinfo{year}{2023}\natexlab{}.
\newblock \showarticletitle{Llm-planner: Few-shot grounded planning for embodied agents with large language models}. In \bibinfo{booktitle}{\emph{Proceedings of the IEEE/CVF International Conference on Computer Vision}}. \bibinfo{pages}{2998--3009}.
\newblock


\bibitem[Sutton(2018)]%
        {sutton2018reinforcement}
\bibfield{author}{\bibinfo{person}{Richard~S Sutton}.} \bibinfo{year}{2018}\natexlab{}.
\newblock \showarticletitle{Reinforcement learning: An introduction}.
\newblock \bibinfo{journal}{\emph{A Bradford Book}} (\bibinfo{year}{2018}).
\newblock


\bibitem[Taffler and Tuckett(2010)]%
        {taffler2010emotional}
\bibfield{author}{\bibinfo{person}{Richard~J Taffler} {and} \bibinfo{person}{David~A Tuckett}.} \bibinfo{year}{2010}\natexlab{}.
\newblock \showarticletitle{Emotional finance: The role of the unconscious in financial decisions}.
\newblock \bibinfo{journal}{\emph{Behavioral finance: Investors, corporations, and markets}} (\bibinfo{year}{2010}), \bibinfo{pages}{95--112}.
\newblock


\bibitem[Talebirad and Nadiri(2023)]%
        {talebirad2023multi}
\bibfield{author}{\bibinfo{person}{Yashar Talebirad} {and} \bibinfo{person}{Amirhossein Nadiri}.} \bibinfo{year}{2023}\natexlab{}.
\newblock \showarticletitle{Multi-agent collaboration: Harnessing the power of intelligent llm agents}.
\newblock \bibinfo{journal}{\emph{arXiv preprint arXiv:2306.03314}} (\bibinfo{year}{2023}).
\newblock


\bibitem[Wang et~al\mbox{.}(2024)]%
        {wang2024survey}
\bibfield{author}{\bibinfo{person}{Lei Wang}, \bibinfo{person}{Chen Ma}, \bibinfo{person}{Xueyang Feng}, \bibinfo{person}{Zeyu Zhang}, \bibinfo{person}{Hao Yang}, \bibinfo{person}{Jingsen Zhang}, \bibinfo{person}{Zhiyuan Chen}, \bibinfo{person}{Jiakai Tang}, \bibinfo{person}{Xu Chen}, \bibinfo{person}{Yankai Lin}, {et~al\mbox{.}}} \bibinfo{year}{2024}\natexlab{}.
\newblock \showarticletitle{A survey on large language model based autonomous agents}.
\newblock \bibinfo{journal}{\emph{Frontiers of Computer Science}} \bibinfo{volume}{18}, \bibinfo{number}{6} (\bibinfo{year}{2024}), \bibinfo{pages}{186345}.
\newblock


\bibitem[Wu et~al\mbox{.}(2023)]%
        {wu2023autogen}
\bibfield{author}{\bibinfo{person}{Qingyun Wu}, \bibinfo{person}{Gagan Bansal}, \bibinfo{person}{Jieyu Zhang}, \bibinfo{person}{Yiran Wu}, \bibinfo{person}{Shaokun Zhang}, \bibinfo{person}{Erkang Zhu}, \bibinfo{person}{Beibin Li}, \bibinfo{person}{Li Jiang}, \bibinfo{person}{Xiaoyun Zhang}, {and} \bibinfo{person}{Chi Wang}.} \bibinfo{year}{2023}\natexlab{}.
\newblock \showarticletitle{Autogen: Enabling next-gen llm applications via multi-agent conversation framework}.
\newblock \bibinfo{journal}{\emph{arXiv preprint arXiv:2308.08155}} (\bibinfo{year}{2023}).
\newblock


\bibitem[Yang et~al\mbox{.}(2024)]%
        {yang2024finrobot}
\bibfield{author}{\bibinfo{person}{Hongyang Yang}, \bibinfo{person}{Boyu Zhang}, \bibinfo{person}{Neng Wang}, \bibinfo{person}{Cheng Guo}, \bibinfo{person}{Xiaoli Zhang}, \bibinfo{person}{Likun Lin}, \bibinfo{person}{Junlin Wang}, \bibinfo{person}{Tianyu Zhou}, \bibinfo{person}{Mao Guan}, \bibinfo{person}{Runjia Zhang}, {et~al\mbox{.}}} \bibinfo{year}{2024}\natexlab{}.
\newblock \showarticletitle{FinRobot: An Open-Source AI Agent Platform for Financial Applications using Large Language Models}.
\newblock \bibinfo{journal}{\emph{arXiv preprint arXiv:2405.14767}} (\bibinfo{year}{2024}).
\newblock


\bibitem[Yu et~al\mbox{.}(2023)]%
        {yangyang2023finmem}
\bibfield{author}{\bibinfo{person}{Yangyang Yu}, \bibinfo{person}{Haohang Li}, \bibinfo{person}{Zhi Chen}, \bibinfo{person}{Yuechen Jiang}, \bibinfo{person}{Yang Li}, \bibinfo{person}{Denghui Zhang}, \bibinfo{person}{Rong Liu}, \bibinfo{person}{Jordan~W. Suchow}, {and} \bibinfo{person}{Khaldoun Khashanah}.} \bibinfo{year}{2023}\natexlab{}.
\newblock \showarticletitle{FinMem: A Performance-Enhanced LLM Trading Agent with Layered Memory and Character Design}.
\newblock \bibinfo{journal}{\emph{arXiv preprint arXiv:2311.13743}} (\bibinfo{year}{2023}).
\newblock


\bibitem[Zhang et~al\mbox{.}(2024)]%
        {zhang2024multimodal}
\bibfield{author}{\bibinfo{person}{Wentao Zhang}, \bibinfo{person}{Lingxuan Zhao}, \bibinfo{person}{Haochong Xia}, \bibinfo{person}{Shuo Sun}, \bibinfo{person}{Jiaze Sun}, \bibinfo{person}{Molei Qin}, \bibinfo{person}{Xinyi Li}, \bibinfo{person}{Yuqing Zhao}, \bibinfo{person}{Yilei Zhao}, \bibinfo{person}{Xinyu Cai}, {et~al\mbox{.}}} \bibinfo{year}{2024}\natexlab{}.
\newblock \showarticletitle{A Multimodal Foundation Agent for Financial Trading: Tool-Augmented, Diversified, and Generalist}. In \bibinfo{booktitle}{\emph{Proceedings of the 30th ACM SIGKDD Conference on Knowledge Discovery and Data Mining}}. \bibinfo{pages}{4314--4325}.
\newblock


\end{thebibliography}

\end{document}